\documentclass[%
 preprint,
 amsmath,amssymb,
 aps,
 pre,
]{revtex4-1}

\usepackage{natbib}
\usepackage{graphicx}
\usepackage{dcolumn}
\usepackage{bm}

\usepackage{color}
\usepackage[normalem]{ulem}

\newcommand{\beq}{\begin{equation}}
\newcommand{\eeq}{\end{equation}}
\newcommand{\beqa}{\begin{eqnarray}}
\newcommand{\eeqa}{\end{eqnarray}}
\newcommand{\ret}{Re_{T0}}

\begin{document}


\title{Notes on the Onset of Clustering in Gas-Solid HCS}

\author{William D. Fullmer}
\affiliation{National Energy Technology Laboratory, Morgantown, WV 26507, USA}
\affiliation{AECOM, Morgantown, WV 26507, USA}

\author{Xiaoqi Li}
\author{Xiaolong Yin} 
\affiliation{Petroleum Engineering Department, Colorado School of Mines, Golden, CO 80401, USA}

\author{Christine M. Hrenya}
\email{hrenya@colorado.edu}
\affiliation{Department of Chemical and Biological Engineering, University of Colorado, Boulder, CO 80309, USA}

\date{\today}

\begin{abstract}
This study contributes to the body of work on instabilities in the homogeneous cooling system focusing on clustering in the multiphase gas-particle system. The critical system size for the onset of instability, $L^*_c$, is studied via three different numerical methods: {\emph i}) particle resolved direct numerical simulation; \emph{ii}) computational fluid dynamics-discrete element method; and \emph{iii}) a two-fluid model derived from kinetic theory. In general, the $L^*_c$ results at several concentrations, inelasticities and initial thermal Reynolds numbers are in good qualitative agreement with one another. Additionally, most of the expected trends (i.e., general $L^*_c (\phi)$ behavior) are observed. However, there is a larger level of quantitative discrepancy between the continuum and discrete particle methods than observed previous (simpler) granular results. While the level of agreement may be expected to decrease with the increased physical complexity of the gas-solid system, a significant time-dependence is revealed and shown to be responsible for some of the oddities in the numerical data. 
\end{abstract}


\maketitle


\section{\label{sec.intro}Introduction}
Owing to its simplicity, the homogeneous cooling system (HCS) is one of the most widely studied particulate systems. In the HCS, particles are uniformly, randomly distributed with approximately normal random velocity components (approximately Maxwellian speed). No external forces act on the system nor body forces on the particles.  The particles simply cool (lose kinetic energy) through dissipative collisions and, in this case, viscous dissipation of the interstitial fluid. The system is unbounded in the sense that it is modeled as boundary free with full periodicity. However, the HCS is not truly unbounded in the sense that the periodic domain size, $L^*$, imposes a length scale that has a significant impact on the dynamics of the HCS \cite{fullmer17a}. 

Haff \cite{haff83} first derived an analytical solution--which now bears his name--for the decay of fluctuating kinetic energy or ``granular temperature'' when a granular (no interstitial fluid) system is in a homogenous cooling state (HCSt), i.e., stable. If the system is sufficiently small in size, the HCS is well described by Haff's cooling law, which has been used to replace the Maxwell-Boltzmann weight function in the derivation of kinetic-theory (KT) based continuum models \cite{garzo07}; verify the KT transport coefficients \cite{brey99, montanero02}; and test code accuracy \citet{fullmer17e}. However, the HCS does not always remain in the (stable) HCSt \cite{fullmer17a}. For a sufficiently large domain size, which depends on the properties of the system, the velocity field may become correlated causing a breakdown in the HCSt decay rate through the generation of localized regions of mean flow or shear fields. This instability is referred to as a velocity-vortex or momentum-mode instability and is characterized by the presence of alternating bands of particle motion. At larger domain sizes, the HCS can further develop concentration inhomogeneities known as the clustering or mass-mode instability. Because the velocity vortex instability develops from the HCSt, linear stability analyses of KT continuum models can be reliably compared to discrete particle simulation results for the onset (in domain size, $L^*$) of instability \cite{mitrano11}. The clustering instability, on the other hand, develops from a nonuniform state, i.e., velocity vortex state \cite{brey99, brilliantov}, implying that a linear-based instability prediction for the onset of clustering may not be valid. Indeed, Mitrano et~al.~\cite{mitrano11} found that linear stability analysis of a frictionless, granular KT continuum model showed a nontrivial discrepancy for clustering instability critical system size, $L^*_c$, when compared to discrete particle simulation data. However, they further showed the clustering onset was well predicted by direct simulation of the KT continuum model (i.e. transient solution of full set of continuum balances) which inherently includes the nonlinear terms of the governing equations. 

For systems with interstitial gas, i.e., the \emph{gas-solid} HCS, analytical solutions for the granular temperature decay rate in the HCSt also exist.  These analytical solutions were validated through direct numerical simulations (DNS) of elastic \cite{wylie00} and inelastic \cite{yin13} systems. Wylie and Koch \cite{wylie00} also found that viscous dissipation from the fluid can induce clustering in the HCS similar to inelastic collisional dissipation--even in the absence of collisional dissipation. The interplay between these two dissipative mechanisms was later studied by Yin et al. \cite{yin13} in the inelastic gas-solid HCS, finding the additional dissipation of the interstitial gas causes an earlier (in time for a fixed domain size) onset of velocity vortex and clustering instabilities. Linear stability analysis of KT two-fluid model (TFM) later verified that the critical length scale for initial velocity vortex instability also decreases compared to the granular theory, which is supported by limited DNS data \cite{garzo16}. 

The current study aims to bridge the gap between the granular clustering instability work of Mitrano et al. \cite{mitrano11} and gas-solid vortex instability work of Garz{\'o} et al. \cite{garzo16}. Namely, the objective is to compare the critical domain size, $L^*_c$, necessary for the onset of clustering between continuum and discrete particle simulations in the gas-solid HCS. DNS serves the role of molecular dynamics simulation in Mitrano et al. \cite{mitrano11} and provides validation data for the comparison. Since the velocity vortex instability precedes the clustering instability, the continuum $L^*_c$ is determined from direct simulation of the full KT-TFM governing equations (as opposed to linear stability analysis). Additionally, a hybrid computational fluid dynamics-discrete element method (CFD-DEM) is also considered in the present work which bridges the DNS and KT-TFM methods. CFD-DEM captures the individual particle motion (as in DNS), but only resolves the average fluid motion, which is coupled to the particles through a mean drag law (as in KT-TFM). In the following sections we overview the three numerical methods, discuss the criteria used to determine critical stability, compare the results of the three methods at several different conditions and finish with a brief summary and future outlook.

\section{\label{sec.methods}Numerical Methods}
For the sake of brevity, the models used in this work are not provided in depth here. Fortunately, all three models have been extensively detailed in previous works which are referenced below. Each model is outlined generally below. The specific modeling choices used in this work are provided for sub-models with multiple options. It is important to note that the interfacial drag force is considered differently by each model. In DNS, drag is captured implicitly, i.e., it is an output. In CFD-DEM, drag acts on the relative velocity between the local mean gas velocity and the individual particle velocity. Finally, in KT-TFM, \emph{mean} drag acts on the mean relative velocity and thermal drag acts on the granular temperature.

\subsection{\label{sec.dns}DNS}
The DNS method is employed in this work to generate data for CFD-DEM and TFM to compare against. As the highest fidelity scheme, DNS resolves all scales of particle and fluid motion, essentially closure free. In this work we use the DNS code SUSP3D developed by Ladd and coworkers \cite{ladd94a, ladd94b, ladd01}. The motion of each particle is solved using Newton’s law with a force determined by integrating the fluid stresses over the particle surface. Particles are marched in time with the fluid timestep with a hard sphere (molecular dynamics) contact model, i.e., particles that will collide during a given timestep are moved to the point of contact and post-collisional velocities are determined from the pre-collisional relative normal velocity and the particle-particle restitution coefficient, $e$. The fluid phase is solved with the lattice-Boltzmann method (LBM) using the D3Q19 velocity model with a two-relaxation time collision model. The lattice spacing (resolution) is approximately 10 lattice units per particle diameter \cite{yin13}. When the distance between particles is less than approximately half a lattice unit a lubrication force model is applied. \cite{nguyen02}. The lubrication force is singular at contact and a cutoff, $\epsilon$, must also be specified.

\subsection{\label{sec.dem}CFD-DEM} 
An intermediary, CFD-DEM, retains the complete particle scale description of DNS while using a coarser description of the fluid, typically with a CFD grid larger than the particle size (Euler-Lagrange). The CFD-DEM model available in the open source MFiX code (\url{https://mfix.netl.doe.gov/}) is used in this work \cite{garg12,garg12b}. Unlike the SUSP3D code, MFiX-DEM employs a soft-sphere linear spring dashpot collision model \cite{cundall79}. In order to mimic the instantaneous collision of the DNS model (also an assumption of the KT derived TFM), the spring constant is set for every case such that the collision duration time, $\delta t_{coll}$, is at least an order of magnitude smaller than the average time between collisions, $\tau_{coll}$, which is at minimum in the initial state:
\beq 
\min {\tau _{coll}}(t) = {\tau _{coll}}({t = 0}) = \frac{{{d_p}}}{{24\phi \chi }}\sqrt {\frac{\pi }{{{T_0}}}} ,
\label{eq.demdt}
\eeq 
where $d_p$ is the particle diameter, $\phi$ is the mean particle (solids) concentration, $\chi$ is the radial distribution function at contact, here using the model of Carnahan and Starling \cite{carnahan69}, and $T_0 = T(t = 0)$, is the initial granular temperature. Then, the spring constant, $k_n$, is set by
\beq 
{k_n} = {m_{eff}}\left( {{\pi ^2} + {{\ln }^2}e} \right)/\delta t_{coll}^2 ,
\label{eq.demkn}
\eeq 
where $m_{eff}$ is the effective mass of the colliding particles ($m_{eff} = m/2$ for monodisperse particles) and $\delta t_{coll} = \tau_{coll}(t_0)/10$. The fluid and solids timesteps used for the numerical integration are set to $dt_{cfd} = \delta t_{coll}$ and $dt_{dem} = \delta t_{coll}/20$, respectively. These variables could be relaxed as the system cools, i.e., set based on $\delta t_{coll}(t)$ rather than $\delta t_{coll}(t_0)$, a scheme which has been successfully tested, however, $\delta t_{coll}$ and its dependents are constant for all CFD-DEM simulations in this work.

\subsection{\label{sec.tfm}KT-TFM}
The KT-TFM represents one additional level of averaging from CFD-DEM in which the particles are treated as the second continuous ``solids'' phase of the TFM (Euler-Euler). A majority of the closures, specifically those related to the solids phase, e.g., solids pressure, viscosity, etc., are derived from KT analogous to (although significantly more complicated than) deriving the Navier-Stokes transport equations from the Boltzmann equation. In this work, we use the KT-TFM of Garz{\'o} et~al.~\cite{garzo16}, alternatively referred to as the GTSH model after the authors. Unlike other granular KT models, the GTSH model was derived specifically for gas-solid flows starting from the Chapman-Enskog equation. The effect of the gas-phase is primarily decomposed into three forces i) mean drag proportional to the difference in mean velocities closed with the model of Beetstra et~al.~\cite{beetstra07}; ii) the thermal drag proportional to the granular temperature (fluctuating kinetic energy of the solids phase) closed with the model of Wylie et~al.~\cite{wylie03}; and iii) the neighbor effect, a stochastic contribution of the unresolved gas velocity field, closed with the model of Koch and Sangani~\cite{koch99}. The details of the models' derivation and a complete listing of equations can be found elsewhere \cite{garzo12, fullmer17b}.

The MFiX code (\url{https://mfix.netl.doe.gov/}) is also used for the numerical solution of the GTSH model. The MFiX TFM employs a finite volume discretization on a staggered grid with first-order upwinding for convective terms. Time advancement is semi-implicit using a SIMPLE-type algorithm with a variable timestep. A relatively fine, cubic grid of side $d_p$ is used in all cases and the system sizes are restricted to integer values. 

\section{\label{sec.setup}Setup}
Modeling the HCS with discrete particle methods (DNS and CFD-DEM) is straightforward: a given number of particles are randomly placed in the domain, the three velocity components are drawn from a normal random distribution, and then scaled to give the desired initial granular temperature with zero mean velocity. Flux renormalization is also carried out periodically in all three numerical methods to prevent drift. The initial fluid velocity is zero everywhere. The discrete initial conditions are uniform in a statistical sense yet, compared to a continuum model, the discrete nature of the system contains an inherent initial perturbation which is a challenge to represent exactly in the KT-TFM model. At present, the best method that we have found is to exactly reproduce the discrete particle initialization procedure, and then filter the discrete information onto the continuum grid using a Gaussian filter with a width of four particle diameters. This procedure allows the initial perturbation to be as consistent as possible between the solution methods, which is important since we will specifically compare solutions at fixed dimensionless times. 

The granular temperature in the discrete particle system is simply taken as the average of the global particle velocity variance 
\beq 
T = \frac{1}{{3{N_p}}}\sum\limits_{i = 1}^{{N_p}} {{{\left| {{{\bf{v}}_i}} \right|}^2}} , 
\label{eq.T}
\eeq 
where mean particle velocities have been dropped because they are set to zero. We note briefly that Eq.~(\ref{eq.T})  only corresponds to the granular temperature in KT-TFM while the system remains in the HCSt so there is no correlated contribution to the fluctuating kinetic energy budget \cite{fullmer18b}. However, this is not a critical distinction in this work since we will not use deviation from the HCSt \cite{mitrano11} as an indicator for the onset of instability. For the discrete particle methods, DNS and CFD-DEM, we follow several previous works \cite{mitrano11, mitrano14, garzo16} that have successfully used the momentum and mass (density) spectra originally studied by Goldhirsch et~al.~\cite{goldhirsch93}. Specifically, we study the ratio of the first and second modes of the momentum and mass spectra, $P_1 / P_2$ and $R_1 / R_2$, for the onset of velocity vortex and clustering instabilities, respectively. If the HCS remains in the HCSt, both spectra increase monotonically so that the ratio of consecutive modes in the spectra are always less than unity. For systems in velocity vortex and/or clustered states, the spectra will have a local maximum in the first mode. Therefore, we take $P_1 / P_2 = 1$ and $R_1 / R_2 = 1$ as the critical stability conditions. For clustering, we also monitor $R_2 / R_3$, however, for all cases studied here $R_2 / R_3  \ge 1$ is preceded by $R_1 / R_2 \ge 1$. There is a slight difference in how the mass mode is used to determine critical stability between DNS and CFD-DEM. Owing to the computational expense, in DNS five replicates of each case are simulated for $t^* = 200$ and if one replicate has attained $R_1 / R_2 \ge 1$, the system is considered unstable. For CFD-DEM which is more computationally affordable, ten replicates are simulated and the mean ${\bar R}_1 / {\bar R}_2$ is considered by averaging the ten replicates at every timestep. A system is considered unstable if the average mass mode becomes unstable at or before $t^* = 200$. Here the dimensionless time $t^*$ is defined as $t^* = t \sqrt{T_0} / d_p$.

Due to the difference between how the mass mode is used to distinguish between stable and unstable systems in DNS and CFD-DEM (i.e., single replicate $R_1 / R_2 > 1$ versus mean ${\bar R}_1 / {\bar R}_2 > 1$), the critical stability point and, possibly more importantly, the associated ``error bars’’ are treated as follows. For DNS, the smallest system with at least one unstable replicate ($R_1 / R_2 > 1$ at $t^* = 200$) is taken as $L^*_c$, the upper error bar is set equal to the lower error bar, which is difference between $L^*_c$ and the largest stable system (no replicates with $R_1 / R_2$). In CFD-DEM, lower and upper bounds (error bars) of the critical system size are taken as the largest stable system (${\bar R}_1 / {\bar R}_2 < 1$ at $t^* = 200$) and the smallest unstable system (${\bar R}_1 / {\bar R}_2 < 1$ at $t^* = 200$), respectively, and $L^*_c$ is simply their algebraic mean. 

Unfortunately, there is not an exact continuum analogue to consider for critical stability in the KT-TFM simulations. Fortunately, however, previous work \cite{mitrano14} comparing critical system sizes for clustering in the inelastic granular HCS between continuum and discrete particle simulation methods found good agreement using a L∞-norm of the concentration field, $\Delta {\phi _{\max }} = \left( {\max {\phi _{i,j,k}} - \min {\phi _{i,j,k}}} \right) / \phi $. The criterion for clustering instability is that $\Delta \phi _{\max }$  be greater than 1\% and increasing at $t^* = 200$. $L^*_c$ is given by the algebraic mean of the largest stable and smallest stable systems. 

The conditions of each HCSt are characterized by a set of four non-dimensional variables: the mean solids concentration, $\phi$; the restitution coefficient, $e$; the density ratio, $\rho^* = \rho_s / \rho_f$; and the initial thermal Reynolds number: $\ret = \rho_f d_p \sqrt{T_0} / \mu_f$. In this work, a constant density ratio of $\rho^* = 1000$ typical to gas-solids multiphase flows. All cases and three numerical methods simulate cubic domains characterized by the side length, $L^* = L/d_p$, with periodic boundary conditions in all three directions.

\section{\label{sec.results}Results and Discussion}
\begin{figure}[t]
\centering\includegraphics[width=0.45\linewidth]{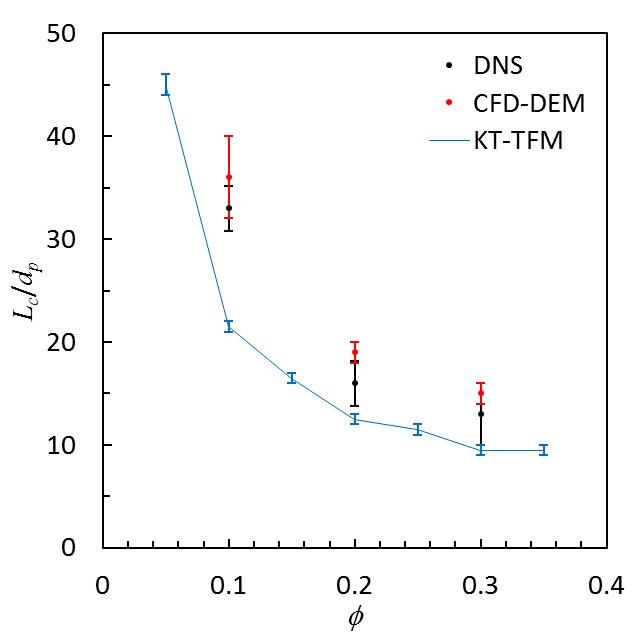}
\centering\includegraphics[width=0.45\linewidth]{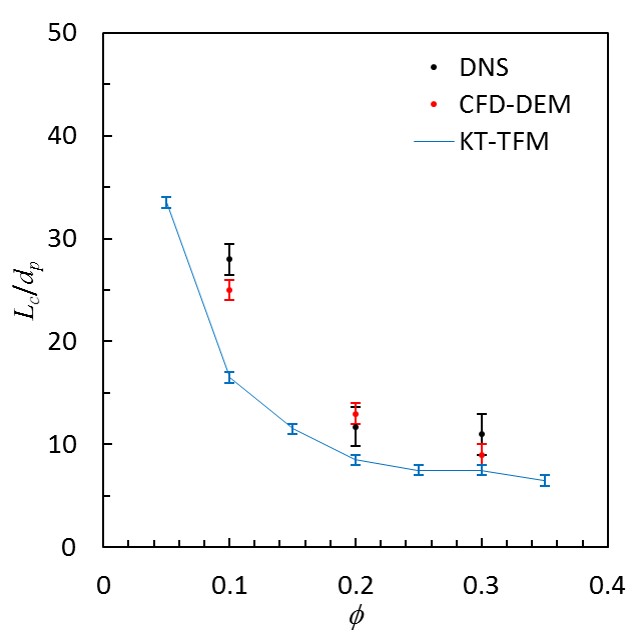}
\label{fig.1}
\caption{Critical dimension for clustering instability in gas-solid HCS as a function of concentration at $\ret = 5$ and $e = 0.9$ (left) or $e = 0.8$ (right).}
\end{figure}

The primary focus of this work is to determine the critical system size, $L^*_c$, necessary for clustering using the aforementioned stability criteria. The critical cubic system size necessary for the onset of clustering instability in the gas-solid HCS is shown in Fig.~\ref{fig.1} as predicted by three different models: DNS, CFD-DEM and KT-TFM (GTSH). The overall trends are consistent with those previously seen from granular HCS results \cite{mitrano11, mitrano14}. Clustering is driven by particle dissipation that increases with increasing concentration (collision frequency), thereby reducing the length scale needed for instability \cite{garzo05, fullmer17b}. Increasing particle inelasticity (decreasing $e$) also increases the dissipation causing a general shift downwards in $L^*_c$ moving from left to right in Fig.~\ref{fig.1} ($e$ from 0.9 to 0.8). Although the overall trends agree, the specific influences of $\phi$ and $e$ on $L^*_c$ are not equivalent among the three models. Most notably, the KT-TFM under-predicts the critical system size predicted by both discrete particle methods at low concentration. Additionally, both KT-TFM and CFD-DEM seem to over-predict the dependence on $e$ relative to DNS. Some discussion regarding the relatively large bounds for the CFD-DEM result at the condition $\phi = 0.1$, $e = 0.9$ and $\ret = 5$ will be revisited later in this section

\begin{figure}[t]
\centering\includegraphics[width=0.45\linewidth]{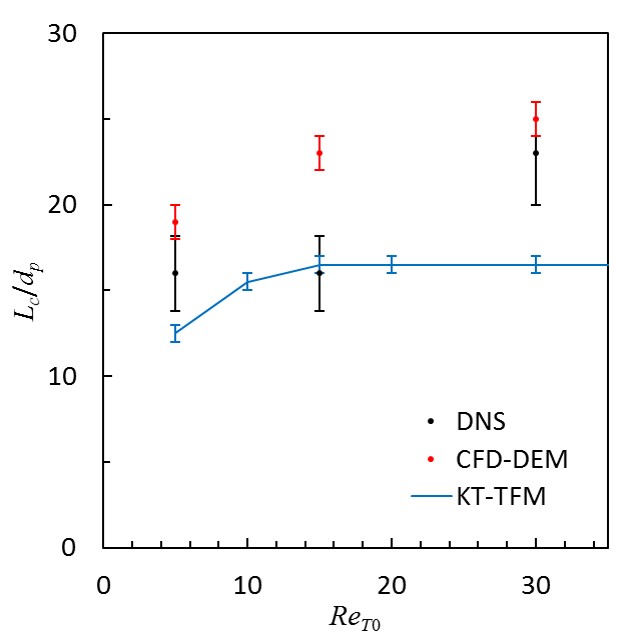}
\centering\includegraphics[width=0.45\linewidth]{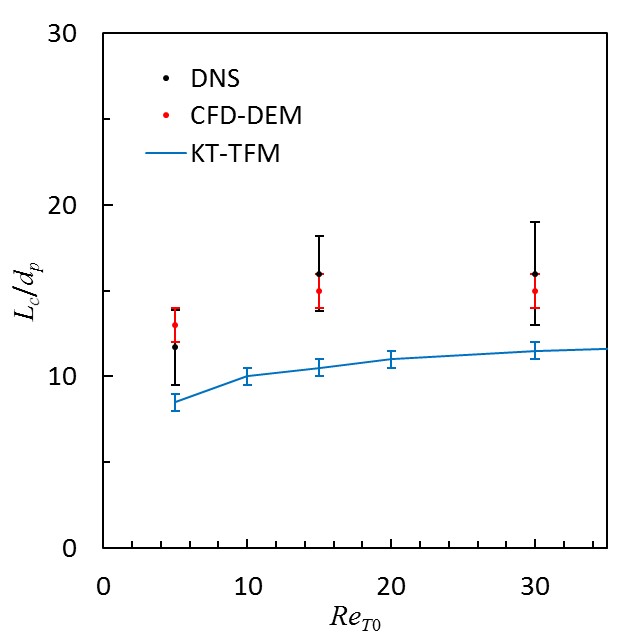} \\
\centering\includegraphics[width=0.45\linewidth]{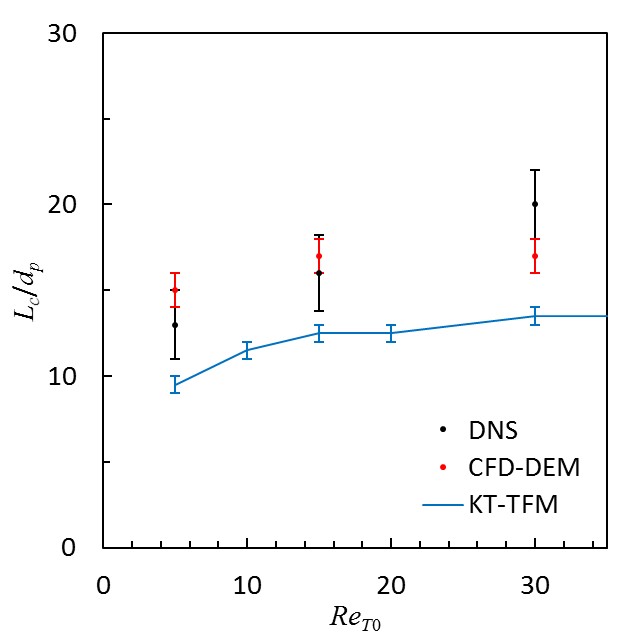}
\centering\includegraphics[width=0.45\linewidth]{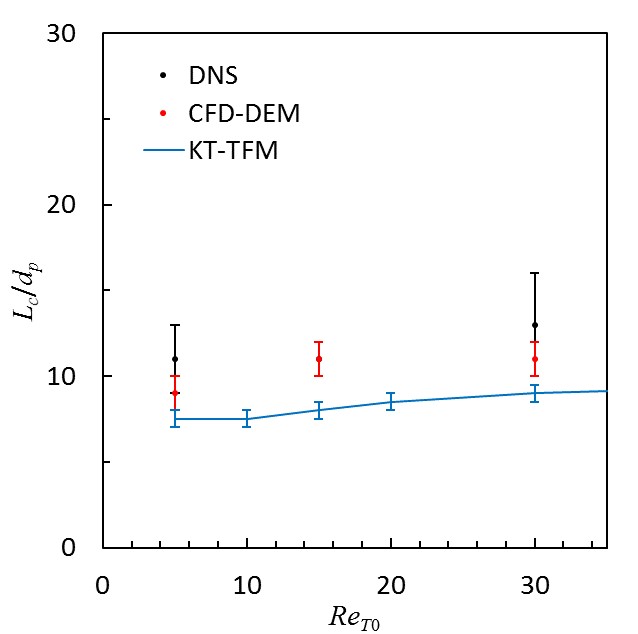}
\label{fig.2}
\caption{Critical dimension for clustering instability in gas-solid HCS as a function of initial thermal Reynolds number, $\ret$, for $\phi = 0.2$ (top row) or $\phi = 0.3$ (bottom row) and $e = 0.9$ (left) or $e = 0.8$ (right).}
\end{figure}

Figure~\ref{fig.2} gives the variation in $L^*_c$ with $\ret$ for two concentrations and two degrees of inelasticity. Results presented in Fig.~\ref{fig.2} show several interesting features. Most importantly, all three methods again correctly predict the most basic trends: $L^*_c$ decreases with decreasing $e$ and $\ret$ and decreases with increasing $\phi$. In the limit of $\ret \to \infty$, the granular HCS is approached as viscous forces become insignificant relative to particle inertia. In terms of stability, increasing $\ret$ increases the time scale over which total dissipation is predominately inelastic (collisional) so that the gas-solid $L^*_c$ should increase asymptotically to the granular $L^*_c$. In Fig.~\ref{fig.2}, the KT-TFM displays this expected behavior results at $\ret = 30$ are comparable to their granular KT analogue \cite{mitrano14}. The CFD-DEM and DNS results, however, are actually larger than their corresponding granular analogue, i.e., the molecular dynamics simulation results \cite{mitrano14}. Superficially, this seems to be an error; the additional source of dissipation from the gas should decrease $L^*_c$ compared to the granular results. The time-dependence of this system must be explored to understand this apparent inconsistency. 

\begin{figure}[t]
\centering\includegraphics[width=0.45\linewidth]{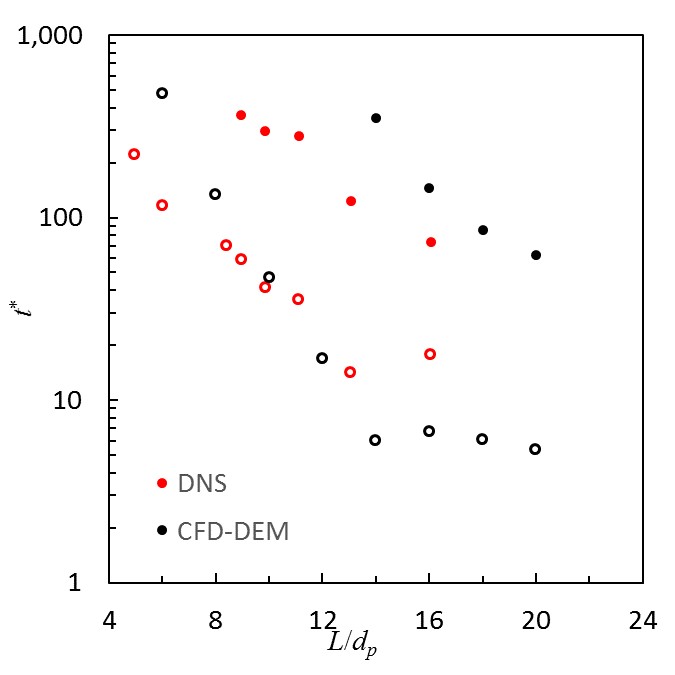}
\label{fig.3}
\caption{Time of the onset of velocity vortex, $t^*_{vv}$ (open points), and clustering, $t^*_c$ (filled points) instabilities for DNS and CFD-DEM at $\ret = 5$, $\phi = 0.3$ and $e = 0.9$}
\end{figure}

While the general trends among the three numerical models are in qualitative agreement with each other, there is undoubtedly some quantitative discrepancy between the models and (for the discrete particle models) their relation to previous granular simulations. The disagreement could be attributed to the assumptions and closures employed by the models. However, we would like to point out that part of this discrepancy is also due to the complexity of determining $L^*_c$ in a time range which is computationally affordable for all simulation methods. Although not the primary focus of this work, the momentum mode ratio $P_1 / P_2$ was also monitored which first revealed a significant $t^*$-dependence in onset of velocity vortex instability, $L^*_{vv}$. To highlight the  $L^*_{vv}(t^*)$ and $L^*_{c}(t^*)$ behavior, the conditions: $\ret = 5$, $\phi = 0.3$ and $e = 0.9$ are extended slightly in the $L^*$-$t^*$ parameter space. The critical times for the onset of velocity vortex $t^*_{vv}$ and clustering $t^*_c$ instabilities for this case are given in Fig.~\ref{fig.3}. For small $L^*$, the velocity vortex instability onset times decrease roughly exponentially (linearly in the semi-log scale of Fig.~\ref{fig.3}) with increasing $L^*$. From the behavior displayed in Fig.~\ref{fig.3} it is rather apparent that the cause of the larger-than-granular $L^*_c$ in Fig.~\ref{fig.2} is due to the simulation time ($t^* = 200$), which is short relative to the $t^* = 5 \times 10^7$ granular simulations \cite{mitrano14}.

Figure~\ref{fig.3} also demonstrates that the shrinking time-to-instability with increasing system size is halted at a certain point and the critical times begin to saturate. The leveling out of $t^*_{vv}$ for the conditions of Fig.~\ref{fig.3} is rather apparent while it appears to just be beginning for $t^*_c$. Such behavior suggests that in gas-solid HCS, the formation of system-spanning structures (vortex or cluster) in large systems may involve multiple time scales that may be associated with the seeding, growth, and coalescence of smaller features. The saturation level, i.e., the fastest time instabilities are reached, was observed to be condition dependent, particularly with respect to concentration. 

Finally, now we are in a position to discuss the relatively large bounds for the low concentration CFD-DEM result in Fig.~\ref{fig.1}. For the condition $\phi  = 0.1$, $e = 0.9$, and $\ret = 5$, the CFD-DEM method did not indicate clustering even up to $L^* = 40$ when using the stability criteria exactly as prescribed in Sec.~\ref{sec.setup}. The critical clustering times for the four largest (CFD-DEM) cases simulated at these conditions are ($L^*$, $t^*_c$): (28, 493), (32, 323) (36, 255), (40, 443). Similar to the behavior observed in Fig.~\ref{fig.3}, $t^*_c$ decreases with increasing system size from 28 to 36, reaches a minimum, and is actually larger at the larger system size of 40. Therefore, we hypothesize that the saturation level (time) may be near $t^* = 200$ for this case. In Fig.~\ref{fig.1}, we take $L^*_c$  = 36 with wide error bars of $\pm$ 4, as it seems reasonable that this system would be considered unstable for slightly different criteria, e.g., if $t^*$ were increased from 200 to 300 or if $R_1 / R_2$ were considered instead of ${\bar R}_1 / {\bar R}_2 = 1$. We further note that system sizes $L^*$ = 32, 36 and 40 all had at least one replicate with $R_1 / R_2 > 1$ at $t^* = 200$ for this condition. 

\begin{table}[!ht]
  \begin{center}
    \caption{Frozen times, $t^*_{\infty}$ approximated from the HCSt.}
    \label{t.tfrozen}
    \begin{tabular}{crcccrccc}
      \hline
      $\phi$ &   &  & $e = 0.9$ &   &  &   & $e = 0.8$ &   \\
          & $\ret =$ & 5 & 10 & 30 & $\ret =$ & 	5 & 10 & 30	 \\												
      0.1	 &  & 	353	 & 	621	 & 	885	 &  & 	296	 & 	519	 & 	738	 \\
      0.2	 &  & 	223	 & 	318	 & 	555	 &  & 	185	 & 	323	 & 	458	 \\
      0.3	 &  & 	142	 & 	202	 & 	353	 &  & 	117	 & 	323	 & 	289	 \\
      \hline
    \end{tabular}
  \end{center}
\end{table}

Unlike the granular HCS which, in theory, never reaches a motionless state, the gas-solid HCS will reach a frozen state in a finite time. In the absence of collisional dissipation (i.e., for particles that have undergone their last collision), a particle moving at a characteristic speed $v_c$ will only travel a finite distance a given an infinite amount of time given by $v_c \tau_f$, where $\tau_f = m/3 \pi \mu_f d_p F^*$ is the viscous relaxation time of the fluid and $F^*$ is the Stokes-deviation of the mean drag law. The gas-solid HCS will essentially freeze, on average, when the viscous length scale associated with the mean particle becomes smaller than the mean free path, $\lambda = d_p / 6 \sqrt{2} \phi \chi$, where $\chi$ is the radial distribution function at contact. A rough estimate of this condition can be approximated by considering the thermal (most probable) speed of the HCSt, i.e., $v_c = c = \sqrt{2 T}$. Then, the condition $c \tau_f / \lambda = 1$ can be re-arranged into the form: 
\beq
Re_{T \infty} = 3 F^* / 2 \phi \chi \rho^* . 
\label{eq.retfrozen}
\eeq
Equation~(\ref{eq.retfrozen}) is of course only approximate as the velocity vortex instability will cause the most probable speed and the mean free path to deviate substantially from the HCSt \cite{yin13}. However, $Re_{T \infty}$ is useful to illustrate how time influences different conditions. Table~\ref{t.tfrozen} provides $t^*_\infty$, the time to reach $Re_{T \infty}$ from $Re_{T0}$ for the conditions of Figs.~\ref{fig.1} and \ref{fig.2}. Although these are only approximate values, $t^*_\infty$ is seen to vary appreciably over the conditions and time domain of interest. 

The previous discussion suggests that, in the absence of infinite computational resources, there is a need for a more robust criterion for the onset of instabilities in the gas-solid HCS which is not so strongly coupled to the simulation time. We note further that the inclusion of an interstitial fluid also significantly complicates the linear stability compared to the granular analogue \cite{garzo16}. In fact, the gas-solid linear stability analysis indicates that all system sizes will eventually become unstable in the limit $t^* \to \infty$ which was attributed to the dual-time scale (collisional and viscous) of the gas-solid HCS. Yet, detecting instability from simulations (DNS, CFD-DEM, KT-TFM) requires that a finite amplitude of inhomogeneity is reached in a finite time. Such a need is particularly important for gas-solid HCS, because viscous relaxation would dissipate the motions of all particles in a finite time. Hence, the time needed to grow an instability to a finite size is limited for gas-solid HCS. For future studies, comparisons may be more fruitful if other stability criteria and characteristics can be explored, such as rate of growth or dissipation of instability, as opposed to velocity or concentration fields selected at (rather arbitrarily) chosen times as attempted in this study.

\section{\label{sec.end}Summary and Outlook}
This study presented the first results for the onset of clustering in the inelastic gas-solid homogeneous cooling system (HCS), building on previous results for the onset of velocity vortex instability in the gas-solid HCS \cite{garzo16} and the onset of velocity vortex \cite{mitrano11} and clustering \cite{mitrano14} instabilities in the granular HCS. Three different methods were applied representing three levels of model fidelity: \emph{i}) direct numerical simulation (DNS), \emph{ii}) computational fluid dynamics-discrete element method (CFD-DEM), and \emph{iii}) kinetic theory two-fluid model (KT-TFM), specifically the GTSH model \cite{garzo12}. In addition to the general degree of fidelity and level of resolution differences of the methods, it is also important to note that all three methods treat the gas-solids coupling differently and used three different criteria to identify an unstable system. 

Despite the underlying differences in the three methods, the results for the onset of clustering in the gas-solid HCS are generally in good qualitative agreement between all three. Specifically, all methods reproduce  the expected trends of decreasing $L^*_c$ (more unstable) with increasing: \emph{i}) inelastic dissipation via higher $\phi$, \emph{ii}) inelastic dissipation via lower $e$, and \emph{iii}) viscous dissipation via lower $\ret$. The quantitative agreement among the three methods is acceptable. However, statistical agreement (i.e., overlapping error bars) is typically only observed between two of the three methods at once. Some of the discrepancy appears to be related to the $t^*$-dependence of $L^*_c$. Unfortunately, running all cases studied in this work for the same time as previous granular simulations is computationally intractable. 

CFD-DEM offers a more computationally affordable discrete particle simulation method for future studies in the gas-solid HCS--at least as an exploratory tool before future DNS parameter sweeps are considered. Specifically, attention should be paid in future studies to re-evaluating the critical stability criteria, improving upon $L^*_c(t^*)$. Previous studies in the granular HCS effectively removed any $t^*$-dependence by running simulations for very long times. Since gas-solid HCS has a frozen state and running simulations for a very long time is not computationally affordable for high fidelity gas-solid simulations, such criteria need to be revisited.

\section*{Acknowledgments}
The authors are grateful to S. Subramaniam, S. Benyahia and J.E. Galvin for insightful discussions on the hydrodynamic closures of the GTSH model and its numerical implementation in MFiX. WDF and CMH would like to acknowledge the funding provided by the National Science Foundation, Grant CBET-1236157. XL and XY would like to acknowledge the funding provided by the National Science Foundation, Grant CBET-1236490.  A portion of this technical effort was performed in support of the National Energy Technology Laboratory’s ongoing research under the RES contract DE-FE0004000. 

This project was funded by the Department of Energy, National Energy Technology Laboratory, an agency of the United States Government, through a support contract with AECOM. Neither the United States Government nor any agency thereof, nor any of their employees, nor AECOM, nor any of their employees, makes any warranty, expressed or implied, or assumes any legal liability or responsibility for the accuracy, completeness, or usefulness of any information, apparatus, product, or process disclosed, or represents that its use would not infringe privately owned rights. Reference herein to any specific commercial product, process, or service by trade name, trademark, manufacturer, or otherwise, does not necessarily constitute or imply its endorsement, recommendation, or favoring by the United States Government or any agency thereof. The views and opinions of authors expressed herein do not necessarily state or reflect those of the United States Government or any agency thereof.

\bibliography{all.bib}

\end{document}